# Phase-Field Modelling of Transformation Pathways and Microstructural Evolution in Multi-Principal Element Alloys

Kamalnath Kadirvel[a#], Zachary Kloenne[a], Jacob K Jensen[a], Hamish Fraser[a], Yunzhi Wang[a]*

[a] Department of Materials Science and Engineering, The Ohio State University, Columbus, Ohio, United States of America.

* Corresponding author, Email: wang.363@osu.edu

# Email: kadirvel.1@buckeyemail.osu.edu

**Abstract:** *The recently developed refractory multi-principle element alloy (MPEA), AlMo$_{0.5}$NbTa$_{0.5}$TiZr, shows an interesting microstructure with cuboidal precipitates of a disordered phase (β, bcc) coherently embedded in an ordered phase (β', B2) matrix, unlike the conventional Ni-based superalloys where the ordered phase (γ', L1$_2$) is the precipitate phase and the disordered phase (γ, fcc) is the matrix phase. It becomes critical to understand the phase transformation pathway (PTP) leading to this microstructure in order to tailor the microstructure for specific engineering applications. In this study, we first propose a possible PTP leading to the microstructure and employ the phase-field method to simulate microstructural evolution along the PTP. We then explore possible PTPs and materials parameters that lead to an inverted microstructure with the ordered phase being the precipitate phase and the disordered phase being the matrix phase, a microstructure similar to those observed in Ni-based superalloys. We find that in order to maintain the precipitates as highly discrete particles along these PTPs, the volume fraction of the precipitate phase needs to be smaller than that of the matrix phase and the elastic stiffness of the precipitate phase should be higher than that of the matrix phase.*

Multi principal element alloys (MPEAs), generally known as high entropy alloys (HEAs), has gained tremendous attentions in the past two decades owing to its potentially better functional, electrochemical and mechanical properties [1-3]. Since single phase MPEAs are not suitable for high temperature structural applications, a significant amount of recent efforts has been devoted to the development of multiphase MPEAs [4-9]. In this study, we focus on a two-phase MPEA, AlMo$_{0.5}$NbTa$_{0.5}$TiZr, that has some exceptional physical and mechanical properties superior to Ni-based superalloys such as low density and high high-temperature yield strength [10]. The microstructure of the alloy was initially characterized to have two β (bcc) phases [11]. However, Jensen et al. [10] found that the matrix (topologically continuous) phase is actually an ordered β′ (B2) phase and the β precipitate phase forms a self-organized, periodic array of cuboidal particles aligned along the <001> directions. At the solutionizing temperature (1400°C) only the $β$ (bcc) phase is present in the microstructure [12]. This microstructure is uncommon because most precipitation-hardened alloys have disordered solution phases as matrix phases and ordered intermetallic phases as precipitate phases. For example, in Ni-base and Co-base superalloys, the disordered ($γ$, fcc) phase is the matrix phase while the ordered phase ($γ′$, L1$_2$) is the precipitate phase. The reason why we expect the matrix phase to be disordered is that at the solutionizing temperature a single disordered phase is present and the ordered phase forms upon cooling. It is also interesting to note that the unique microstructural features observed in AlMo$_{0.5}$NbTa$_{0.5}$TiZr, including (i) narrow particle size distribution and (ii) highly regular and aligned cuboidal particles arrays, resemble well those of the $γ + γ′$ microstructure in Ni-base and Co-base superalloys. The formation of the $γ + γ′$ microstructures have been explained and modeled via the conventional nucleation and growth mechanism, i.e., the $γ′$ phase precipitates out of a supersaturated parent $γ$ phase solid solution. It should be noted that under specific heat treatment conditions (such as rapid-cooling), it is possible to obtain $γ + γ′$ two-phase microstructures through a spinodal-mediated phase separation pathway [13, 14]. However, what lead to the "inverted superalloy-like" β′ + β microstructure observed in AlMo$_{0.5}$NbTa$_{0.5}$TiZr is still unclear. For a similar alloy system (Al$_{0.5}$NbTa$_{0.8}$Ti$_{1.5}$V$_{0.2}$Zr), two possible PTPs mediated through spinodal decomposition [15] have been proposed, which could lead to $β′ + β$ microstructure. In order to tailor and optimize the microstructure for specific engineering applications, it becomes critical to understand the PTP leading to this β′ + β microstructure. In general, materials with an ordered phase as the continuous matrix phase may not have the well-balanced properties that the superalloys can offer and, thus, there is a need to invert the β′ (matrix) + β (precipitate) microstructure into a β (matrix) + β′ (precipitate) microstructure by adjusting the alloy composition and heat treatment schedule. Experimental efforts are on-going in this area to tailor the β′ + β microstructures in MPEAs [9, 15-18].

In this work, using a prototype model system and phase-field simulations, we demonstrate a possible PTP leading to the β′ + β microstructure and study effects of the equilibrium volume fractions of the β′ and β phases and their elastic properties on inverting the β′ + β microstructure into a β + β′ one. The proposed PTP consists of (i) congruent ordering of β into β′ upon cooling, (ii) spinodal decomposition of β′ into solute-lean and solute-rich B2 phases, and (iii) congruent disordering of the solute-lean β′ phase into β, leading to a β′ + β microstructure.

Our prototype binary system has two phases: (i) a disordered phase that represents the bcc ($β$) phase (ii) an ordered phase that represents the B2 ($β′$) phase. The free energy curves of the B2 ($f^{B2}$) and bcc ($f^{bcc}$) phases as function of solute concentration ($c$) of the prototype model binary system at a temperature within the BCC+B2 field of the phase diagram are shown in Fig. 1. Both the free energies ($f^{B2}, f^{bcc}$) and solution concentration ($c$) are presented in non-dimensional forms. The non-dimensional solute concentration ($c$) can be related to real solute concentration ($x$) using $c = (x - x_{bcc})/(x_{B2} - x_{bcc})$ where $x_{bcc}$ and $x_{B2}$ are the equilibrium solute concentrations of the bcc and B2 phases, respectively. The B2 phase has a miscibility gap and the spinodal region within which, $\partial^2 f^{B2}/\partial c^2 < 0$, is in between the two cross symbols in Fig. 1(a). The dashed line in Fig. 1(a) represents the absolute instability of the bcc phase with respect to spontaneous ordering, i.e., a disorder – order transition without an energy barrier. For both Alloy 1 and Alloy 2 indicated in Fig. 1(a), spinodal decomposition in the B2 phase will take place following congruent ordering of the starting bcc phase. This kind of spinodal decomposition is called conditional spinodal [19, 20] and is known to occur in Al-Fe system [21]. The equilibrium microstructures in both alloys are mixtures of B2 and bcc phases with different volume fractions of these phases. We chose the solute atoms of our prototype system to represent $\beta'$ forming elements such as Al and Zr. Hence, the equilibrium B2 phase is solute-rich ($c = 1$) and the equilibrium bcc phase is solute-lean ($c = 0$).

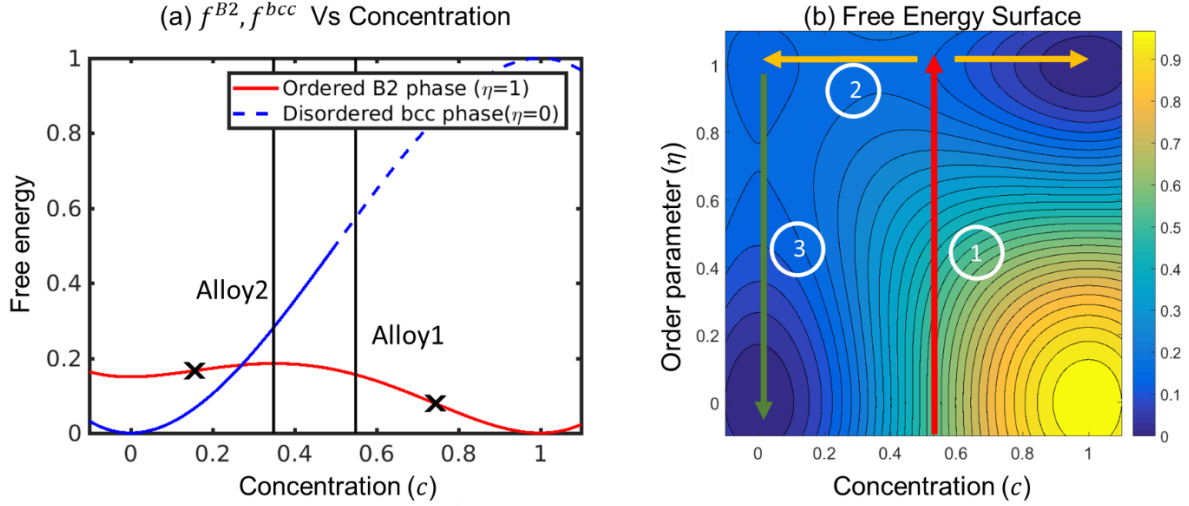

Fig. 1. (a) The free energy of the ordered B2 phase $f^{B2}$ and the free energy of the disordered bcc phase $f^{bcc}$ as a function of composition. Spinodal boundaries of the ordered phase are marked as black cross. The dashed lines in the disordered state represent unstable region. The alloys used in this study as marked as Alloy 1 ($c_{avg} = 0.55$) and Alloy 2 ($c_{avg} = 0.35$). (b) The free energy surface with arrows shows the various steps in the phase transformation: 1: Congruent ordering (**CO**), 2: Spinodal decomposition of the ordered phase (**SD**), and 3: Congruent disordering of the solute-lean region (**CD**).

The phase-field model consists of two order parameters: (i) concentration $c(r,t)$ and (ii) long-range order (lro) parameter $\eta(r,t)$ that distinguishes the disordered bcc phase ($\eta = 0$) from the ordered B2 phase ($\eta = 1$). The total free energy of the system $F[c(r,t), \eta(r,t)]$ is formulated as a functional of these order parameters,

$$F = \int dr \left[ f(c,\eta) + \frac{\kappa_c}{2}|\nabla c|^2 + \frac{\kappa_\eta}{2}|\nabla \eta|^2 \right] + E^{el}[c(r,t), \eta(r,t)] \quad (1)$$

where $f(c,\eta)$ is the local chemical free energy density, $\kappa_c$ is the gradient energy coefficient for concentration, $\kappa_\eta$ is the gradient energy coefficient for the lro parameter and $E^{el}[c(r,t), \eta(r,t)]$ is the elastic strain energy functional of the system. The local free density $f(c,\eta)$ is approximated by

$$f(c,\eta) = f^{bcc}(1 - h(\eta)) + f^{B2}h(\eta) + \omega g(\eta) \quad (2)$$

where $h(\eta) = \eta^2(3 - 2\eta)$ is an interpolation function, $g(\eta) = \eta^2(1-\eta)^2$ is a double-well function and $\omega$ is the barrier height between the ordered and disordered phases. A contour plot of the free energy surface, $f(c,\eta)$, is shown in Fig. 1(b) and its projection onto the $f$-$c$ plane is shown in Fig. 1(a). The color bar in the Fig. 1(b) represents the magnitude of the local free energy density, $f(c,\eta)$.

For simplicity, we assume that the lattice misfit is only a functional of concentration following the Vegard law [22-24]. The elastic energy functional is given by

$$E^{el} = \int dr \left[ \frac{1}{2} C_{ijkl}(\epsilon_{ij} - \epsilon_{ij}^T)(\epsilon_{kl} - \epsilon_{kl}^T) \right] \quad (3)$$

where $C_{ijkl}$ is the elastic modulus field, $\epsilon_{kl}^T$ is the transformation strain field and $\epsilon_{ij}$ is the total strain field. Both the $C_{ijkl}$ and $\epsilon_{kl}^T$ are functions of concentration $c(\mathbf{r},t)$:

$$C_{ijkl}(c) = C_{ijkl}^{bcc} \times (1-c) + C_{ijkl}^{B2} \times (c) \qquad (4)$$

$$\epsilon_{ij}^T = \epsilon^{oo} c \delta_{ij} \qquad (5)$$

where $\epsilon^{oo} = \frac{1}{a}\frac{da}{dc}$, $a$ is the lattice parameter, $C_{ijkl}^{bcc}$ is the modulus of the disordered phase and $C_{ijkl}^{B2}$ is the modulus of the ordered phase. The total strain field $\epsilon(r)$ is determined by the mechanical equilibrium equation: $\nabla \cdot \sigma = 0$. When the modulus of the system is homogeneous ($C_{ijkl}^{bcc} = C_{ijkl}^{B2}$), the elastic energy $E^{el}$ can be directly written as an explicit functional of the order parameters as derived by Khachaturyan et al. [25, 26] using the Green's function solution in the reciprocal space. This closed form expression of $E^{el}$ greatly increases the numerical efficiency of the model. When the modulus of the system is inhomogeneous, an iterative procedure [27, 28] is used to compute the strain field.

The time-evolution of the concentration field, $c(\mathbf{r},t)$, and the lro order parameter field, $\eta(\mathbf{r},t)$, follows the Cahn-Hilliard equation [29] and Allen-Cahn equation [30] (also known as time-dependent Ginsburg-Landau equation), respectively,

$$\frac{\partial c}{\partial t} = \nabla \left( M \nabla \left( \frac{\delta F}{\delta c} \right) \right) + \zeta_c \qquad (6)$$

$$\frac{\partial \eta}{\partial t} = -L \frac{\delta F}{\delta \eta} + \zeta_\eta \qquad (7)$$

where, $M$ is the chemical mobility, $L$ is the kinetic coefficient characterizing the order-disorder transition, $\zeta_c$ and $\zeta_\eta$ are the Langevin-noise terms for concentration and lro parameter, respectively. The Langevin-noise terms are added to the governing equations in order to incorporate thermal fluctuations in our model [31, 32]. The governing equations (Eq.6 and Eq.7) are solved using the semi-implicit spectral method [33]. When the disordered or ordered phase is metastable, the congruent disorder ↔ order transition occurs by nucleation, which is simulated by the Langevin noise term $\zeta_\eta$ in the Allen-Cahn (Eq.7) equation [34, 35]. For simplicity, the Langevin noise for concentration is chosen to be zero, $\zeta_c = 0$. All the numerical values of the model parameters used in the simulation are listed in Table 1. All simulations are performed in two dimensions with a system size of $512 l_o \times 512 l_o$ where $l_o$ is the grid size. The length scale in the phase field model is determined by the ratio of the interfacial energy to the elastic energy. Assuming an interfacial energy of $\gamma = 50 mJ/m^2$ for the coherent interface between the B2 and bcc phases and a typical shear modulus ($C_{44}$) of 182GPa for the bcc phase, the grid size $l_o$ is determined to be 4.2 $nm$.

The simulation results obtained for Alloy 1 is shown in Fig. 2, along with the PTP marked on the free-energy contour. In this case, we assumed homogeneous modulus ($C_{ijkl}^{bcc} = C_{ijkl}^{B2}$) and a lattice misfit of 0.2% between the B2 and bcc phases. The evolution of the system in Fig. 2 is represented through three columns where the first column shows the concentration field $c(\mathbf{r},t)$, the second column shows the schematic PTP on the free energy surface where the current state of the system is marked by the white open squares, and the third column show the order parameter field $\eta(\mathbf{r},t)$. The figure shows eight snapshots of the microstructural evolution process marked as (1-8) below the PTP in the second column. The initial condition of the simulation (snapshot (1)) is a homogeneous bcc phase corresponding to the state of the system at the homogenization temperature before cooling.

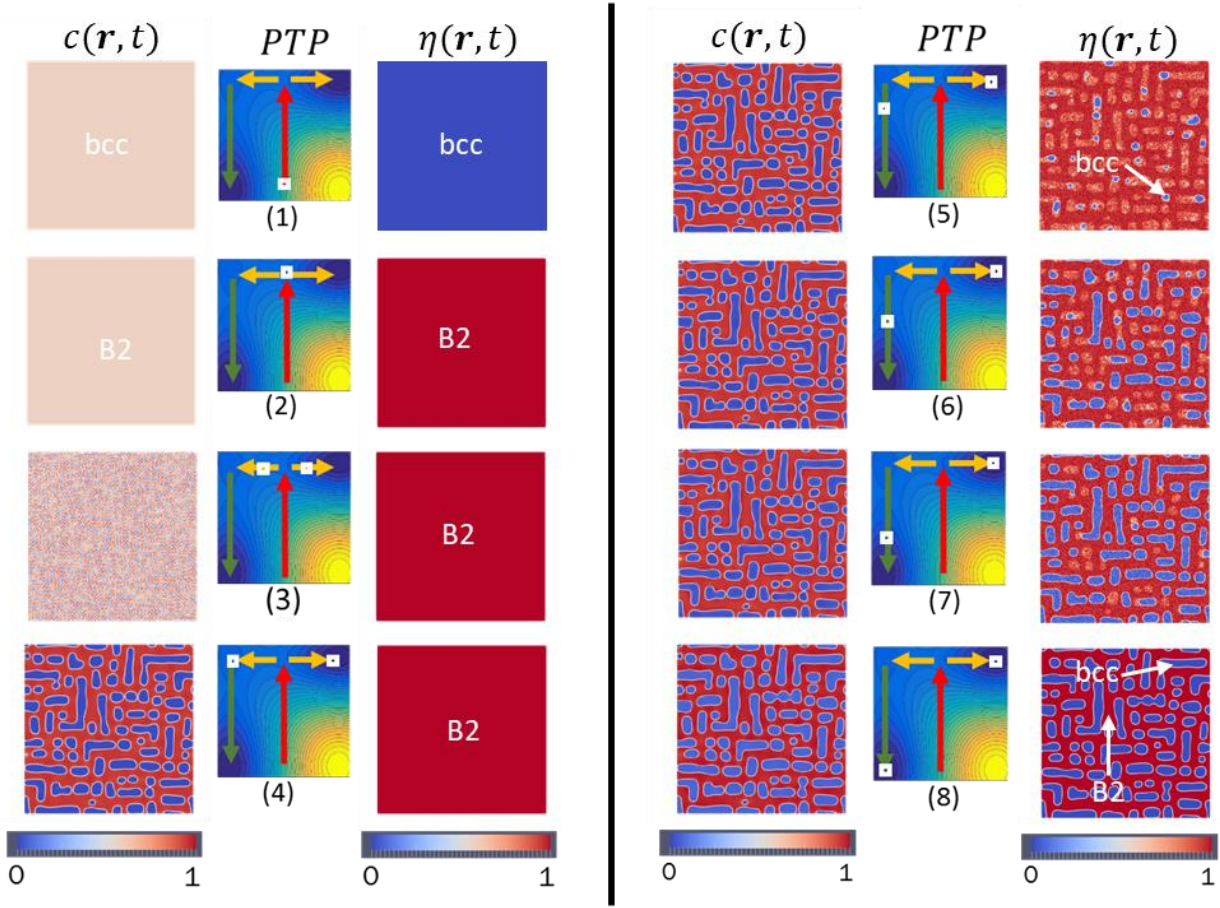

Fig. 2. Microstructural evolution in Alloy 1 represented by the concentration field $c(\mathbf{r},t)$ in the 1st column, phase transformation pathway PTP in the 2nd column, and the lro parameter field $\eta(\mathbf{r},t)$ in the 3rd column at different stages of the phase transformations: (1) Before congruent ordering; (2) After congruent ordering; (3) During spinodal decomposition; (4) Post spinodal decomposition; (5) Nucleation of disordered phase; (6) & (7) During congruent disordering; 8) Final B2 (matrix) +bcc (precipitates) microstructure.

The snapshot (2) shows the congruently ordered (B2) state formed during isothermal aging at a lower temperature, where $\eta(\mathbf{r},t) = 1$ everywhere in the system without any concentration modulations. Since the alloy composition is located in the absolutely unstable region (shown as blue dashed line in Fig. 1(a)) for Alloy 1, the congruent ordering is a barrierless transformation and the transformation can proceed without the assistance of the Langevin noise term (i.e., $\zeta_\eta(\mathbf{r},t) = 0$ ). For Alloy 2, the alloy composition is located in the metastable region (shown as blue solid line in Fig. 1a) and then the Langevin noise term, $\zeta_\eta(\mathbf{r},t)$, is required to initiate the congruent ordering reaction. But in both cases, the microstructure will reach the congruently ordered (B2) single-phase state first before spinodal decomposition. Note that the B2 phase has two anti-phase domains and can form anti-phase domain boundaries (APB) during the congruent ordering. We assume that the anti-phase domain size is much larger than our simulated system size and ignore the presence of APBs. This assumption is justified from the experiments where the size of the anti-phase domains is much larger than the size of the disordered precipitates [10].

The snapshots (3) and (4) capture the spinodal decomposition process in the B2 phase. The early stage of the spinodal decomposition shown in snapshot (3) resembles a basket-weave microstructure with two orthogonal concentration waves along the elastically soft directions (i.e., <001>) superimposed on top of each other. After the concentration modulation develops completely, we observe ordered solute-lean domains embedded in an ordered solute-rich matrix, aligned along the <001> elastically soft directions as per the $C_{ijkl}$ chosen in our model. The solute-lean ordered region is metastable, which can be seen from the free energy surface (Fig. 1(b)) and undergo congruent disordering. The snapshots (5), (6), (7) and (8) show the congruent disordering process. In the solute-lean B2 region, the bcc particles are formed by nucleation and growth (assisted by thermal fluctuations in the lro parameter that is captured by the Langevin noise term in the simulations), as indicated by the blue patches (snapshot (5), the right column) in the $\eta(r,t)$ field. Finally, most of the solute-lean regions become disordered domains embedded in a continuous, solute-rich ordered B2 phase matrix, as shown in snapshot (8).

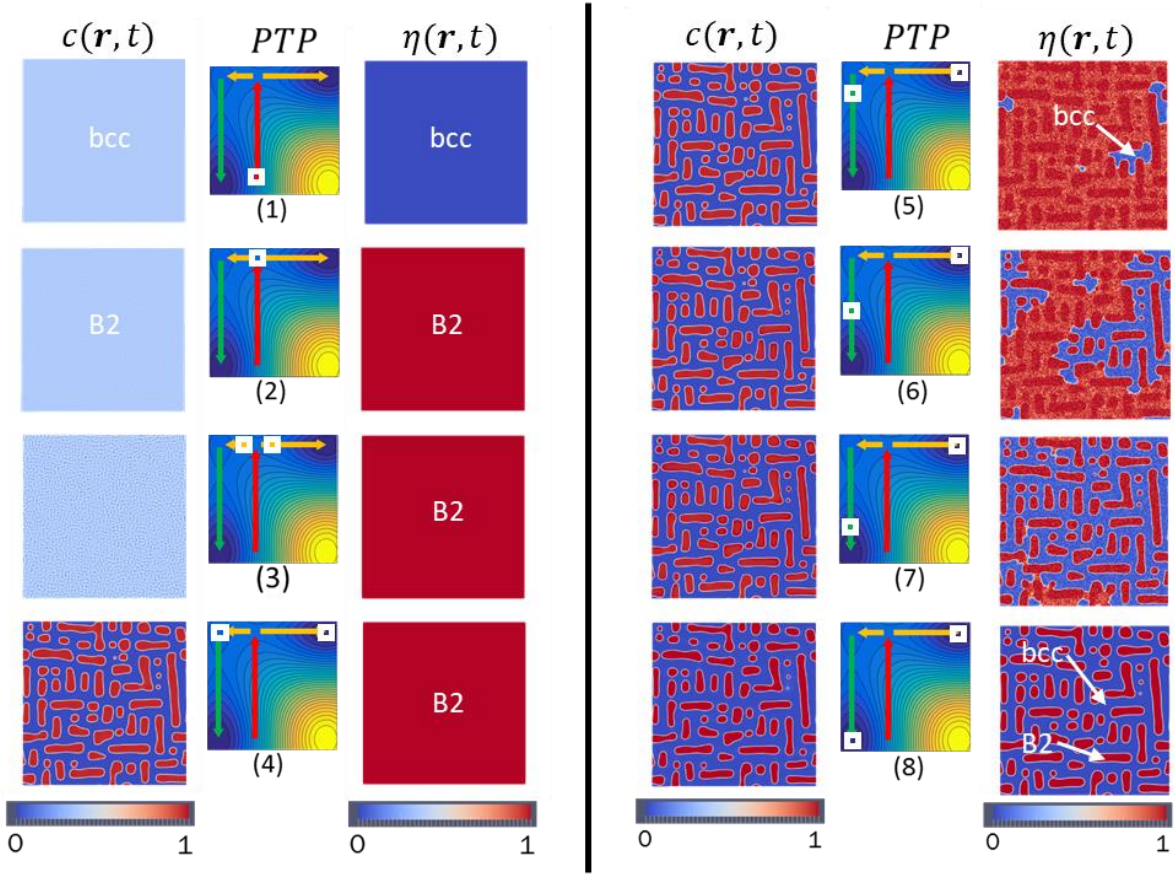

Fig. 3. Microstructural evolution in Alloy 2 represented by the concentration field $c(r,t)$ in the 1st column, phase transformation pathway PTP in the 2nd column, and the lro parameter field $\eta(r,t)$ in the 3rd column at different stages of the phase transformations: (1) Before congruent ordering; (2) After congruent ordering; (3) Early stage spinodal decomposition; (4) Post spinodal decomposition; (5) Nucleation of disordered phase; (6) & (7) During congruent disordering; 8) Final bcc (matrix) +B2 (precipitates) microstructure.

Thus, our proposed PTP leads to a microstructure where discrete bcc precipitates embedded in a continuous B2 matrix, which agrees with that observed in the experiment [10]. The comparison between the microstructure of the annealed sample and the phase-field simulated microstructure is shown in Fig.B1 (Supplementary Material). We have also simulated the microstructural evolution in Alloy 2 with parameter given in the Table 1. The microstructural evolution in Alloy 2 leads to the inverted microstructure $\beta$(matrix) + $\beta'$(precipitate) as shown in the Fig. 3.

From the above simulation results (Fig. 2 and Fig. 3), we realize that whether the ordered or disordered phase forms the continuous matrix phase is determined by the spinodal decomposition step of the PTP ( marked in Fig. 1b). At the end of the congruent ordering step, the system is a homogeneous single

B2 phase without any concentration modulations. In the congruent disordering step (snapshots (5)-(8) in both Fig. 2 and Fig. 3), only the solute-lean B2 regions can transform to bcc phase and the morphology of the solute-lean and solute rich regions do not change. Hence the morphologies (i.e., discrete, continuous, or bi-continuous) of the ordered and disordered phases is predetermined by the spinodal decomposition process.

A clear understanding of the phase transformation pathway (PTP) is essential to design the alloy composition and heat-treatment, which will help us to tailor the microstructure for specific engineering applications. Below we show an example of tailoring the microstructure under our proposed PTP by modifying the alloy composition and elastic modulus.

The as-cooled sample from the experiments has discrete precipitates consisting of cuboidal particles and small square shaped platelets, well-aligned in the <001> direction, which are in contrast to the elongated plate-like (rod in 2D) precipitates observed in our simulation (Fig. 2). A more quantitative definition of the discreteness can be found in the Supplementary material (Section A). This could be due to the modulus mismatch between the two phases. We further studied the effect of modulus mismatch and alloy composition (i.e., volume fractions of the two phases) on microstructural evolution during spinodal decomposition. It was shown in the literature that an elastically more compliant (i.e., soft) phase tend to wrap around an elastically stiffer (i.e., hard) phase in order to minimize the elastic energy during phase separation [36, 37]. In a complex multicomponent system such as our MPEA, we do not have reliable measurements of the elastic modulus $C_{ijkl}$ of the individual phases, namely the solute-lean B2 (which will congruently disorder into bcc phase in the final microstructure) and solute rich B2 phases, to identify the elastically soft phase. Hence, we consider two extreme cases: (i) Case-A: soft solute-rich phase $R_{elas} = C_{ijkl}^{B2}/C_{ijkl}^{bcc}$=0.6; (ii) Case-B: hard solute-rich phase $R_{elas} = 1.4$. In order to obtain the desired $R_{elas}$, the modulus of the solute-lean B2 phase was kept constant (values given in Table 1), while changing the modulus of the solute-rich B2 phase. We simulated the spinodal decomposition process in these two cases for both Alloy 1 and Alloy 2. Note that the equilibrium volume fraction of the solute-rich phase in Alloy 1 (~60%) is higher than Alloy 2 (~40%). The simulated microstructures are shown in Fig. 4. Even though these microstructures are similar to those obtained by Li et al. [38], the modulus mismatch considered in the current study ($R_{elas} = 1.4$ or 0.6) is much larger than that used in Li et al. [28] ($R_{elas} \approx 1.03$ or 0.97). What we find is that such a large modulus mismatch is necessary to achieve significantly increased discreteness of the precipitate phase as compared to those obtained in the homogeneous modulus counterpart. We also performed a systematic study to quantify discreteness of the microstructure as a function of the inhomogeneity ratio $R_{elas}$ (see Supplementary Fig. A1).

In Case A, we observe a highly discrete solute-lean precipitates in Alloy 1, as shown in Fig. 4(a3). As the volume fraction of the solute-lean phase is low in this alloy, it becomes the precipitate phase (i.e. topologically discontinuous phase) and, because the matrix phase is elastically softer, the modulus mismatch further increased the discreteness of the precipitates. In Alloy 2, where the volume fraction of the solute-lean phase is high, it becomes the matrix at the early stage of the decomposition process (Fig. 4 (b2)). As the decomposition proceeds, however, the solute-rich phase becomes more topologically connected because it is elastically soft, leading to a nearly bi-continuous microstructure (Fig. 4(b3)). In Case B, we observe a nearly bi-continuous microstructure in Alloy 1 (Fig. 4(c3)) and discrete solute-lean particles in a solute-rich matrix in Alloy 2 (Fig. 4 (d3)). The formation mechanisms of these microstructures can be explained with similar reasoning as that mentioned in Case A. The microstructure observed in Alloy 2 is interesting as it will give the desired inverted microstructure (β matrix + β′ precipitate) after congruent disordering of the solute-lean phase.

We have noted that the current simulations do not reproduce exactly the microstructures observed in the experiment (see, e.g., Fig. 4 of Ref.[10]), i.e., the presence of both cube-shaped precipitates and plate-shaped satellite precipitates in between. Research is ongoing to identify the underlying mechanisms for the formation of such an interesting microstructure.

Recently, Li et al. [38] developed a phase field model to simulated the B2/bcc microstructures observed in HEAs. Even though, the simulated microstructure resembles the microstructure observed in the experiments, they considered spinodal decomposition into a B2 and a bcc phase, which is not an isostructural transformation. Thus, the actual PTP in these alloys could be similar to what we have considered in the current study, i.e., congruent ordering of bcc into B2, followed by spinodal decomposition of the B2 phase into two B2 phases having different compositions, and one of them congruently disorders into bcc [21], or follow other non-conventional PTPs such as pseudospinodal [39, 40] and mechanochemical spinodal [41] that occur between phases having different crystal structures.

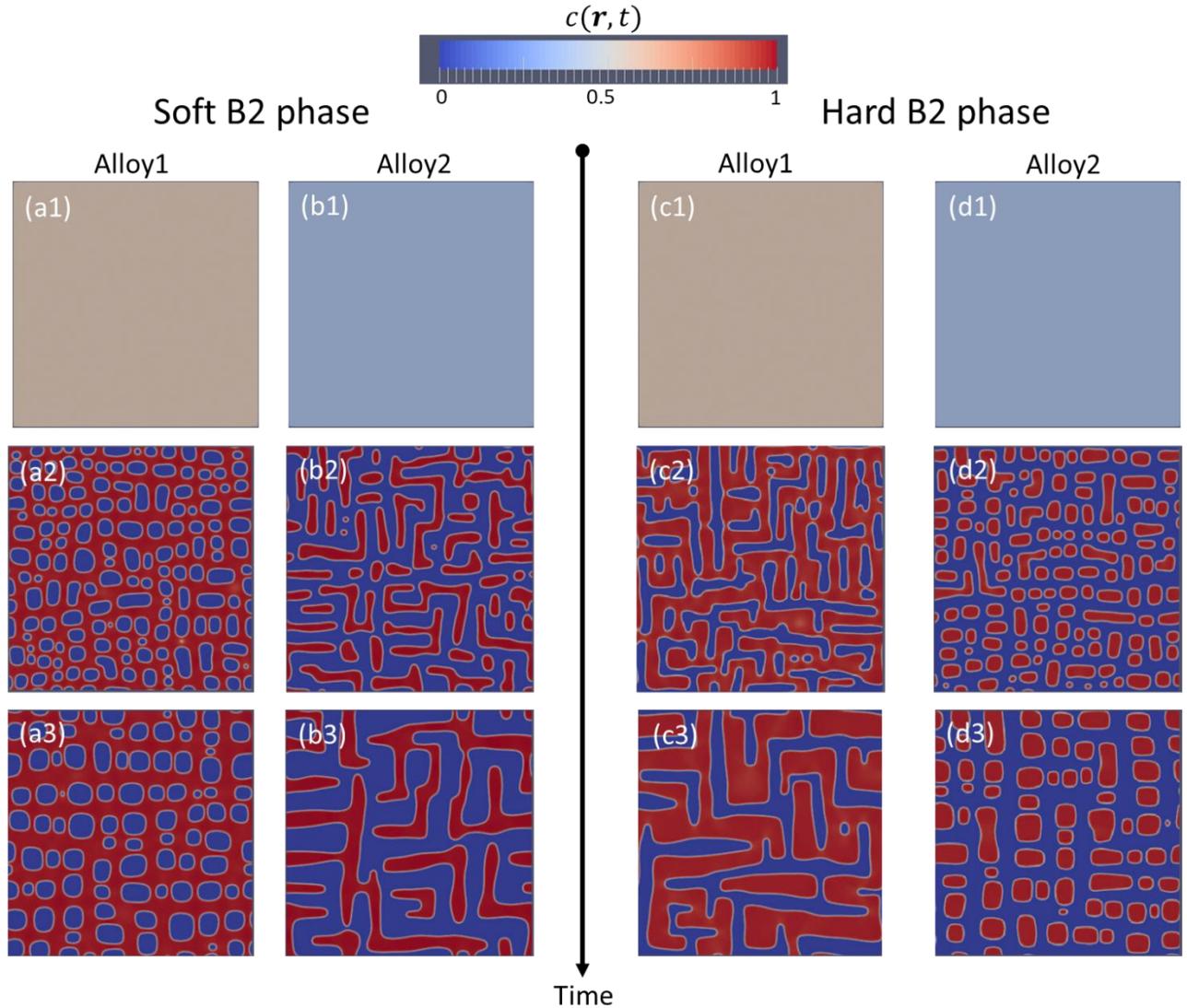

Fig. 4. Microstructural evolution during the SD (Spinodal Decomposition) step for the alloys Alloy 1 and Alloy 2 where different rows represent different time steps with the top row being the initial condition. (a1,a2,a3) Alloy 1 with $R_{elas} = 0.6$. (b1,b2,b3) Alloy 2 with $R_{elas} = 0.6$. (c1,c2,c3) Alloy 1 with $R_{elas} = 1.4$. (d1, d2, d3) Alloy 2 with $R_{elas} = 1.4$.

    In summary, we have proposed and analyzed a phase transformation pathway (PTP) for phase separation in AlMo$_{0.5}$NbTa$_{0.5}$TiZr MPEA. Phase field simulations of microstructural evolutions along this PTP for different alloy compositions and different elastic modulus mismatch between the two co-existing phases were performed. They show that both the $\beta'$(matrix) + $\beta$ (precipitate) microstructure which is observed in experiments and its inverted counterpart $\beta$ (matrix) + $\beta'$ (precipitate) can be obtained from the

proposed PTP. They also show that the modulus mismatch between the phases can be used to increase the discreteness of the precipitate phase.

## SUPPLEMENTARY MATERIAL

See the Supplementary material for the effect of modulus inhomogeneity on the microstructure and the experimental microstructure.

## ACKNOWLEDGEMENT

The authors would like to acknowledge the financial support from Air Force Office of Scientific Research (AFOSR) under grant FA9550-20-1-0015. Computational resources for this work is provided by Ohio Supercomputer Center [42] under the project PAS0971.

## DATA AVAILABILITY

The data that support the findings of this study are available from the corresponding author upon reasonable request.

Table 1. Numerical values (non-dimensional) of the model parameters used in the simulation.

| Parameter type | Parameter name | Symbol | Value |
|---|---|---|---|
| Kinetic parameters | Chemical Mobility | M | 1 |
| | Interface kinetic coefficient | L | 30 |
| Free energy parameters | Gradient energy coefficient for $c(r,t)$ | $\kappa_c$ | 2 |
| | Gradient energy coefficient for $\eta(r,t)$ | $\kappa_\eta$ | 2 |
| | Double-well barrier height | $\omega$ | 1 |
| | Free energy of bcc phase | $f^{bcc} = c^2(3-2c)$ | |
| | Free energy of B2 phase | $f^{B2} = \frac{3}{2}[c^2(1-c)^2 + 0.1(1-3c^2+2c^3)]$ | |
| Elastic energy parameters | Elastic modulus tensor | $C_{11}$ | 32211 |
| | | $C_{12}$ | 21906 |
| | | $C_{44}$ | 15453 |
| | Lattice misfit | $\epsilon^{oo}$ | 0.2% |
| Noise parameters | Gaussian random number generator with zero mean and unit standard deviation | $\rho$ | - |
| | Langevin noise for $\eta(r,t)$ | $\zeta_\eta$ | $\rho\sqrt{\frac{2(k_BT)L}{\Delta t}}$ Where (i) $\Delta t$ is the discretization timestep (iii) $k_BT$ is the non-dimensional thermal energy |
| | $\zeta_\eta$ was applied for a duration of $\Delta t^{langevin}$ in the following cases: (i) Congruent disordering of Alloy 1 ($k_BT = 0.06$; $\Delta t^{langevin} = 45$) (ii) Congruent ordering of Alloy 2 ($k_BT = 0.2$; $\Delta t^{langevin} = 0.2$) (iii) Congruent disordering of Alloy 2 ($k_BT = 0.06$; $\Delta t^{langevin} = 45$) | | |

# Supplementary Material

Section A: Effect of inhomogeneity ratio $R_{elas} = C^{B2}_{ijkl}/C^{bcc}_{ijkl}$ on the microstructure

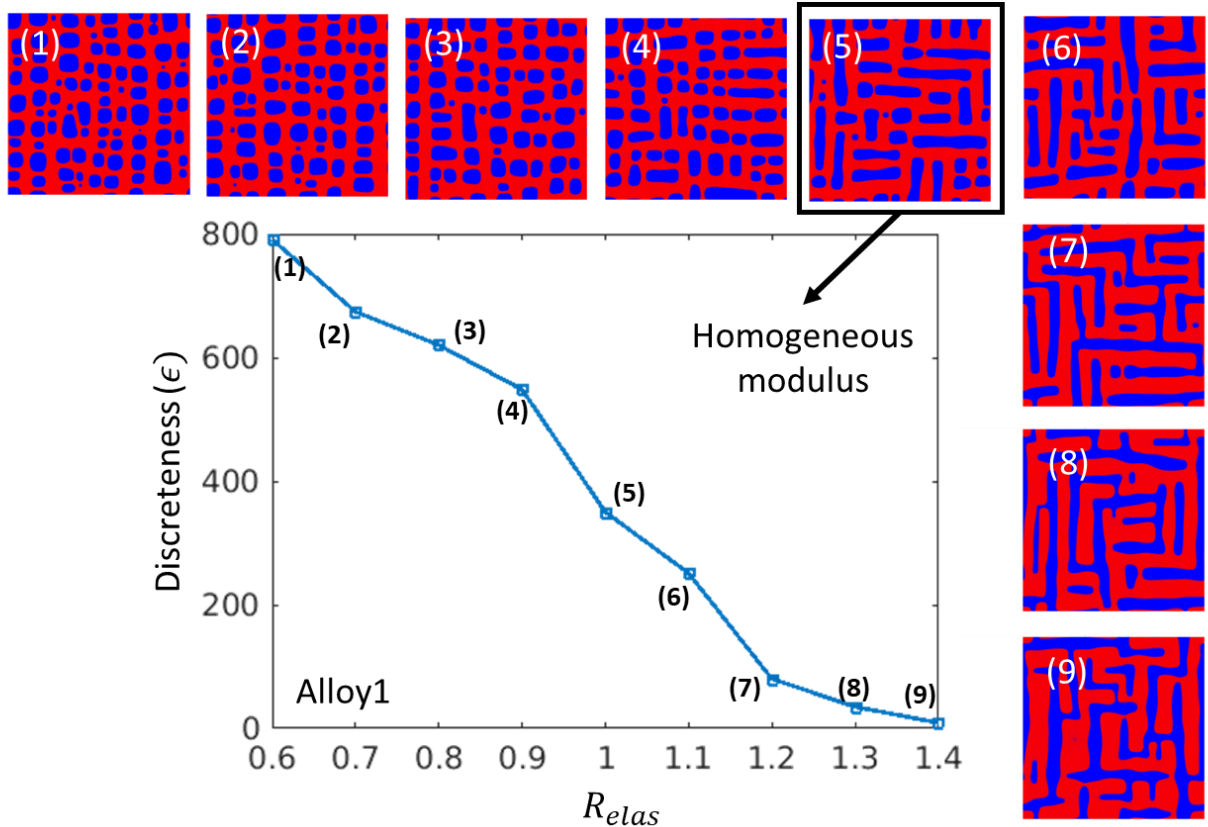

Figure A1: Calculated discreteness ($\epsilon$) as a function of the elastic inhomogeneity ratio $R_{elas}$. Discreteness of the microstructure decreases with increasing $R_{elas}$ for Alloy1.

Modulus inhomogeneity plays a critical role in determining the discreteness of the microstructure. We performed a systematic study by varying the inhomogeneity ratio ($R_{elas}$) similar to the simulation in Fig. 4 of the main text. However, in this study we focused only on Alloy1 and varied $R_{elas}$ from 0.6 to 1.4 in steps of 0.1. In order to quantitatively compare the discreteness of the microstructure, we defined a discreteness parameter

$$\epsilon = |N_B - N_R|$$

where $N_B$ is the number of blue (solute-lean B2) particles and $N_R$ is the number of red (solute-rich B2) particles in the microstructure. The quantities $N_B$ and $N_R$ are calculated from the $3 \times 3$ supercell of the simulated microstructure. The simulated microstructures for Alloy1 along with the discreteness value is showed in Fig.A1. As in Alloy1, the volume of red phase (~60%) is higher than the blue phase, red phase becomes the matrix for the homogeneous modulus case ($R_{elas} = 1.0$) with the discreteness value of about $\epsilon = 350$. As observed from Fig.A1, discreteness $\epsilon$ changes continuously as we vary $R_{elas}$ and the discreteness is $\epsilon \approx 0$ for $R_{elas} = 1.4$. Based on this study we chose $R_{elas}$=1.4 and $R_{elas}$ =0.6 as the extreme values for analysis in Fig. 4 of the main text.

## Section B: Experimental characterization of $AlMo_{0.5}NbTa_{0.5}TiZr$

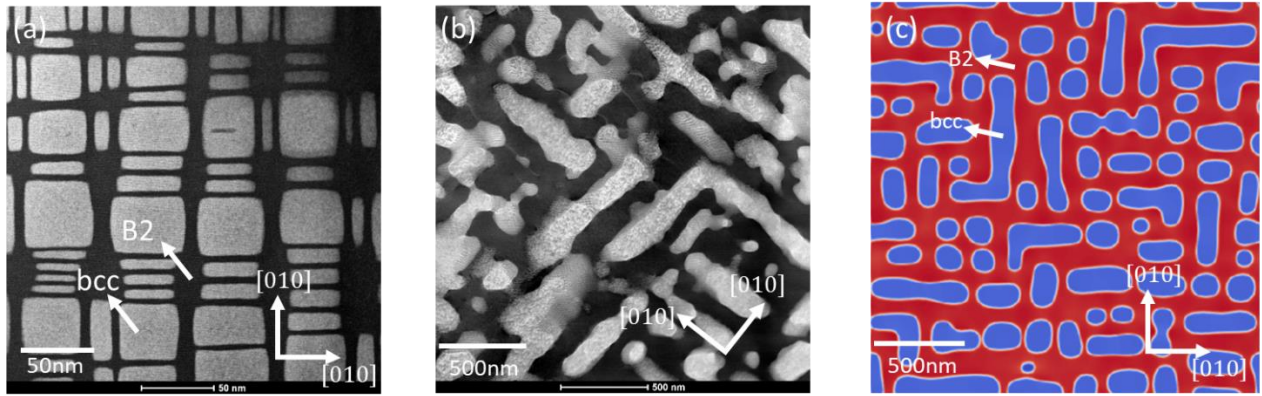

Figure B1: (a) STEM-HAADF image of solutionized and furnace cooled ($10^oC$/min) sample (b) STEM-HAADF image of the sample annealed at $1000^oC$ for 6 hours. (c) The simulated microstructure ($\eta$ −profile) from the phase-field model.

The sample preparation procedure is elaborated in "O.N. Senkov, S.V. Senkova, C. Woodward, Acta Mater. 68 (2014) 214–228" and characterization procedure is detailed in "J.K. Jensen, B.A. Welk, R.E.A. Williams, J.M. Sosa, D.E. Huber, O.N. Senkov, G.B. Viswanathan, H.L. Fraser, Scripta Materialia 121 (2016) 1-4". Detailed results of the experiment will be submitted as another article to the special collection "Metastable High Entropy Alloys" of Applied Physics Letters.